\documentclass[12pt]{article}
\usepackage{graphicx}
\usepackage{amssymb}
\usepackage{bm}

\newcommand{\re}{\mathrm{Re}\,}
\newcommand{\im}{\mathrm{Im}\,}

\newcommand{\opr}[1]{\mathsf{#1}}

\renewcommand{\[}{\begin{equation}}                               %
\renewcommand{\]}{\end{equation}}

\newcommand{\abg}{(\alpha \beta + |\gamma|^2)}

\newcommand{\laplace}{\triangle}

\newcommand{\rest}{D(\opr{h}_{l}^{*})}
\renewcommand{\O}{\mathcal{O}}

\begin{document}


\title{Resonance asymptotics \\ in the generalized Winter model}

\author{Pavel Exner and Martin Fraas}
\date{\small Nuclear Physics Institute, Czech Academy of Sciences,
25068 \v{R}e\v{z} near Prague, \\
Doppler Institute, Czech Technical University, B\v{r}ehov\'{a} 7,
11519 Prague, Czechia }

\maketitle


\begin{flushleft}
{\small We consider a modification of the Winter model describing
a quantum particle in presence of a spherical barrier given by a
fixed generalized point interaction. It is shown that the three
classes of such interactions correspond to three different types
of asymptotic behaviour of resonances of the model at high
energies.}
\end{flushleft}

\vspace{3em}

\noindent Models with contact interactions are popular because
they allow us to study properties of quantum systems in a
framework which makes explicit solutions possible. It was found in
the beginning of the eighties that the usual $\delta$ interaction
on the line has a counterpart, named not quite fortunately
$\delta'$, and a little later the complete four-parameter class of
the generalized point interactions (GPI's) was introduced
\cite{GeK, Se1}. Properties of these interactions are now well
understood -- see \cite{solvable} for a rather complete
bibliography.

The GPI's fall into different classes according to their behaviour
at low and high energies. The most simple manifestation can be
found in scattering. While a $\delta$-type barrier behaves as a
``usual'' regular potential becoming transparent at high energies,
the $\delta'$-like one on the contrary decouples asymptotically in
the same limit. In addition, there is an \emph{intermediate class}
for which both the reflection and transmission amplitudes have
nonzero limits as the energy tends to zero or infinity \cite{EG}.
Furthermore, in Kronig-Penney-type models describing periodic
arrays of such interactions, the indicated classes differ by the
gap behaviour at high energies; this has important consequences
for spectral nature of the corresponding Wannier-Stark systems
\cite{AEL, ADE}.

In this letter we discuss the GPI's in the context of the
generalized \emph{Winter model} describing a quantum particle in
presence of a rotationally symmetric surface interaction supported
by a sphere; we suppose that the interaction is repulsive
corresponding to a barrier which gives rise to tunneling decay and
scattering resonances. Our aim is to show that the said GPI
classes correspond to different high-energy asymptotics of the
resonances, specifically, we are going to show that for the
$\delta$-type interaction the resonance lifetime tends to zero
with the increasing energy, for $\delta'$ interaction it increases
to infinity asymptotically linearly in the resonance index, and
that there is an intermediate case for which the resonance
lifetime still tends to zero but in a slower way than it the
$\delta$ case, $\O(n^{-1})$ vs. $\O((n\ln n)^{-1})$.

Recall that the Winter model was introduced in \cite{winter,
Green}, originally for the $\delta$ barrier in which case the
Hamiltonian can be formally written as
 $$
 H= -\laplace + \alpha \delta(|x|- R)\,,
 $$
where $R$ is the radius of the sphere $S_R$ and $\alpha$ is the
coupling constant. A thorough analysis including the $\delta'$
extension can be found in \cite{AGS}; an extension to multiple
spheres is given in \cite{Shabani} and other related results are
reviewed in \cite{solvable}. From the mathematical point of view
such models are described by means of spherically symmetric
self-adjoint extensions of the symmetric operator $\dot{H}:\:
\dot{H}\psi = -\laplace\psi,$ defined on $D(\dot{H}) = \{f \in
W^{2,\,2} (\mathbb{R}^3 \setminus S_R),\, f(x)=\nabla f(x)=0\quad
\mbox{for}\quad x \in S_R\}$, i.e. the restriction of the free
Hamiltonian to functions vanishing together with their derivatives
on the sphere. These extensions form a four-parameter family
characterized by the conditions (\ref{eq:bound}) below.

The spherical symmetry allows us to reduced the analysis to a
family of halfline problems by partial wave decomposition
\cite{Shabani}). Using the isometry $U\,:\,L^2((0,\,\infty),
r^{2}dr)\, \rightarrow \, L^2(0,\,\infty)$ defined by
$Uf(r)=rf(r)$ we write
 $$
 L^2(R^3) = \bigoplus_l U^{-1}L^2(0,\,\infty)\otimes S_1^{(l)}\,, \quad
 \dot{H} = \bigoplus_l U^{-1}\,\dot{h}_{l}\,U \otimes I_l
 $$
where $I_l $ is the identity operator on $S_1^{(l)}$ and
$\dot{h}_{l} = -\frac{d^2}{d\,r^2} + \frac{l(l+1)}{r^2}$ with the
domain $D(\dot{h}_{l}) = \{f \in \rest;  \, f(R) = f'(R) = 0\}$;
here $\rest=W^{2,\,2}((0,\,\infty) \setminus R)$ if $l\ne 0$ while
for $l=0$ the requirement $f(0+) = 0$ is added. The equation
 $$
 \dot{h}_{l}^*f = k^2f , \quad \im\, k > 0\,,
 $$
has two linearly independent solutions \cite{AGS, Shabani}, namely
 \begin{eqnarray*}
       \Phi^{(1)}_{k,\,l}(r) &=& \left\{ \begin{array}{l}
      \frac{i\pi}{2} \sqrt{r}J_{\nu}(kr)H^{(1)}_{\nu}(kR)\sqrt{R} \\
      \frac{i\pi}{2} \sqrt{R}J_{\nu}(kR)H^{(1)}_{\nu}(kr)\sqrt{r} \\
   \end{array}   \right. \\
   \Phi^{(2)}_{k,l}(r) &=& \left\{
\begin{array}{lc}
   i\frac{\pi}{2}J_\nu(kr)\sqrt{r}(H^{'(1)}_\nu(kR)\sqrt{R}k
   + H^{(1)}_\nu(kR)\frac{1}{2\sqrt{R}}) \\
   i\frac{\pi}{2}(J'_\nu(kR)k\sqrt{R} +
   J_\nu(kR)\frac{1}{2\sqrt{R}})H_\nu^{(1)}(kr)\sqrt{r}
\end{array} \right.
\end{eqnarray*}
for $0<r\leq R$ and $r>R$, respectively, where $\nu = l + 1/2 $;
thus we have a four-parameter family of extensions which can be
parameterized via boundary conditions at the singular point $R$.

There are many different, mutually equivalent forms of such
boundary conditions. We will use those introduced in \cite{EG}: a
(rotationally invariant) surface interaction Hamiltonian is
characterized by
 \begin{equation} \label{eq:bound}
 \begin{array}{l}
   f'(R_+)\! -\! f'(R_-) = \frac{\alpha}{2}\!\left(f(R_+)\!
   +\! f(R_-)\right) + \frac{\gamma}{2}\!(f'(R_+)\! +\! f'(R_-))
   \\ [.3em]
   f(R_+)\!  -\! f(R_-)  = -\frac{\bar\gamma}{2}\!(f(R_+)\!
   + \! f(R_-)) + \frac{\beta}{2}\!(f'(R_+)\! +\! f'(R_-))
 \end{array}
 \end{equation}
for $ \alpha,\,\beta \in \mathbb{R}$ and $\gamma \in \mathbb{C} $;
we will call such an operator $H_{\alpha,\, \beta, \,\gamma}$. The
three GPI classes mentioned in the introduction are the following,
 \begin{itemize}
 \item \emph{$\delta$-type:} $\quad \re\gamma=\beta=0$,
 \vspace{-.5em}
 \item \emph{intermediate type:} $\quad \re\gamma \neq 0,\,\beta=0$,
 \vspace{-.5em}
 \item \emph{$\delta'$-type:} $\quad \beta \neq 0 $.
 \end{itemize}
Notice that the intermediate class contains the
``scale-invariant'' interaction considered recently by
Hej\v{c}\'{\i}k and Cheon \cite{HC}; it corresponds to
$\alpha=\beta=0$ and
 $$
 \gamma= \frac{h-h^{-1}+2i\sin\phi}{h+h^{-1}+2\cos\phi}\,,
 $$
where $h$ is their $\alpha$. Notice also that this interaction is
not as exotic as it might look; it appears in description of a
free quantum motion on a regular metric tree \cite{SoS} where
$\phi=0$ and $h=\sqrt{N}$ where $N$ is the branching number.

Of the other forms of the boundary conditions let us mention those
of \cite{Fulop}. Denote by $F$ the column vector of $f(R+)$ and
$f(R-)$, with $F'$ similarly corresponding to the one-sided
derivatives at $r=R$. Then the self-adjoint extensions are
parametrized by
 \begin{equation} \label{eq:canbou}
 (U - I)F + i(U + I)F' = 0\,,
 \end{equation}
where
 $$
 U = e^{i \xi} \left( \begin{array}{cc}
 u_1 & u_2 \\ -\bar{u}_2 & \bar{u}_1
 \end{array} \right)
 $$
is a $2\times2$ unitary matrix, $u_1,\,u_2 \in \mathbb{C}$ with
$|u_1|^2 + |u_2|^2 =1 $ and $\xi \in [0,\,\pi)$. The boundary
conditions (\ref{eq:canbou}) are related to (\ref{eq:bound}) by
 \begin{eqnarray*}
 u_1 &=& \frac{-2(\alpha + \beta) +4 i \re \gamma}{\sqrt{(\alpha \beta +|\gamma|^2)^2 + 4\alpha^2 +4\beta^2 + 8|\gamma|^2 + 16 }}\\
 u_2 &=&\frac{1}{2i} \frac{(\alpha\beta + |\gamma|^2 -4) - 4 i \im \gamma}{\sqrt{(\alpha \beta +|\gamma|^2)^2 + 4\alpha^2 +4\beta^2 + 8|\gamma|^2 + 16 }}  \\
  \tan \xi &=& \frac{\alpha \beta + |\gamma|^2 +4}{2(\alpha - \beta)} \,,
 \end{eqnarray*}
in particular, the case $u_2 = 0$, or
 $$
 \alpha \beta + |\gamma|^2 = 4 \quad \mbox{and} \quad \im \gamma = 0
 $$
corresponds to the separated motion inside and outside the
barrier; in such a case there is infinitely many embedded
eigenvalues on the positive real axis. The three distinct classes
in the question can also be described through the matrix $U$. For
instance, the $\delta'$-type requires $\re u_1+\cos\xi\ne 0$,
while for the former two types this quantity vanishes, the
$\delta$-type corresponding to $\im u_1=0$. It is easy to cast
these conditions into a more elegant form,
 \begin{itemize}
 \item \emph{$\delta$-type:} $\det (U + I) =0,\, \sigma_1U^T\sigma_1 = U$,
 \vspace{-.5em}
 \item \emph{intermediate type:} $\det (U + I) =0,\, \sigma_1U^T\sigma_1 \neq U$,
 \vspace{-.5em}
 \item \emph{$\delta'$-type:} $|\det (U + I)|>0$.
 \end{itemize}
In particular, the $\delta$ and intermediate type are the family
of point interactions for which $-1$ is an eigenvalue of
corresponding matrix $U$, with the $\delta$-type being the
subfamily invariant under the $\mathcal{PT}$-transformation,
induced by the map $U \to \sigma_1 U^T \sigma_1$, where $\sigma_1$
is the first Pauli matrix \cite{ChFT}.

Another equivalent form of the boundary conditions which can be
traced back to \cite{GeK, Se1} is
 $$
 \left(\begin{array}{c}
 f(R+) \\ f'(R+)
 \end{array} \right) = \Lambda
 \left(\begin{array}{c}
 f(R-) \\ f'(R-)
 \end{array} \right),
 $$
where $\Lambda$ is a matrix of the form
 $$
 \Lambda = e^{i \chi} \left( \begin{array}{cc}
a & b\\
c & d
 \end{array} \right), \quad \chi \in [0,\pi),\quad a,\,b,\,c,\,d \in
\mathbb{R},\quad ad - bc =1\,;
 $$
in terms of parameters used above $\Lambda$ can be written as
 $$
 \frac{1}{\alpha \beta + |\gamma|^2 - 4 + 4 i \im
 \gamma}  \left( \begin{array}{cc}
                    \alpha \beta + |\gamma|^2 + 4 - 4 \re \gamma &
                    4 \beta \\
                    4 \alpha & \alpha \beta + |\gamma|^2 + 4 + 4 \re \gamma
                \end{array} \right).
 $$
Using the transformations
 $$
 f \rightarrow \left\{\begin{array}{cl}
 e^{i\phi} f & r<R \\
 e^{i\theta} f & r \geq R
 \end{array}\right.
 $$
it is easy to see that extension corresponding to $U$ and $U'$ are
unitarily equivalent if $|u_2|=|u'_2|$. Moreover, to any extension
described by $(\alpha',\,\beta',\,\gamma')$ there is a unitarily
equivalent one with $(\alpha,\,\beta,\,\gamma)$ such that $\im
\gamma = 0$. Indeed, the above transformation with $\theta = 0$
means the replacement
 $$
 \Lambda \to \tilde{\Lambda} = e^{i(\chi - \varphi)}
\left( \begin{array}{cc}
            a& b \\
            c&d
        \end{array} \right)\,,
 $$
so in any class of unitarily equivalent extensions there is one
with a real transfer matrix; this occurs if $\im\gamma = 0$.

After these preliminaries let us turn to our main subject. If
$u_2\ne 0$ the Hamiltonian has no embedded eigenvalues. The
singularities do not disappear, however, instead we have an
infinite family of resonances. To find them, we have derive an
explicit expression for the resolvent which can be achieved by
means of Krein's formula \cite[App.~A]{solvable}: we have
 $$
 (H_{(\alpha,\,\beta,\,\gamma),\,l} - k^2)^{-1}
 = (H_{0,\,l} - k^2)^{-1} + \sum_{m,n = 1}^2
 \lambda_{mn}(\alpha,\,\beta,\,\gamma)
    (\Phi^{(n)}_{-\bar{k},\,l}, \cdot) \Phi^{(m)}_{k,\,l}\,,
 $$
where $\Phi^{(m)}_{k,\,l}$ are the solutions to the deficiency
equations given above. The coefficients $\lambda_{i,\,j}$ are
found from the fact that the resolvent maps into the domain of
$H_{(\alpha,\beta, \gamma)}$ which means, in particular, that
$(H_{(\alpha,\,\beta, \,\gamma),\,l} - k^2)^{-1}g$ must for any
$g\in L^2((0,\infty))$ satisfy the boundary conditions
(\ref{eq:bound}). A straightforward computation then gives
 \begin{eqnarray*}
 \lambda_{11} &=& \frac{\alpha - \Phi^{'(2)}_k(R)(\alpha \beta
 + |\gamma|^2)}{\det \lambda}\,,  \\
    \lambda_{12} &=& \frac{\gamma + \Phi^{(2)}_k(\bar{R})(\alpha \beta
    + |\gamma|^2)}{\det \lambda}\,, \\
    \lambda_{21} &=& \frac{\bar\gamma + \Phi^{(2)}_k(\bar{R})(\alpha \beta
    + |\gamma|^2)}{\det \lambda}\,, \\
    \lambda_{22} &=& \frac{ -\beta - \Phi^{(1)}_k(R)(\alpha \beta
    + |\gamma|^2)}{\det \lambda}\,,
 \end{eqnarray*}
where the denominator is given explicitly by
 \begin{equation} \label{det}
\det \lambda = -1 -\alpha \Phi^{(1)}_k(R) + \beta \Phi^{'(2)}_k(R)
- (\gamma + \bar{\gamma})\Phi^{(2)}_k(\bar R) \\ -
\frac{1}{4}(\alpha \beta + |\gamma|^2)
 \end{equation}
and by $\Phi^{(2)}_{k,\,l}(\bar{R}):= \frac12 \left(
\Phi^{(2)}_{k,\,l}(R+) + \Phi^{(2)}_{k,\,l}(R-) \right)$.

As usual the resonances are identified with the poles of the
resolvent at the unphysical sheet, $\im k <0 $, their distance
from the real axis being inversely proportional to the lifetime of
the resonance state. We will discuss the asymptotic in momentum
$k$-plane, more specifically, the behaviour of $\im k_n$
corresponding to the $n$th resonance, given through roots of the
equation
 \begin{equation} \label{eq:pol.eq}
 \det\lambda(k,\,\alpha,\,\beta,\,\gamma) = 0
 \end{equation}
in the open lower halfplane. We are going to demonstrate that
 \begin{itemize}
 \item for the \emph{$\delta$-type} the quantity $\im k_n$
 increases logarithmically w.r.t. the resonance index as $n\to\infty$,
 \vspace{-.5em}
 \item for the \emph{intermediate type} $\im k_n$ is
asymptotically constant,
 \vspace{-.5em}
 \item finally, for the \emph{$\delta'$-type} with $\beta>0$ the
 quantity $\im k_n$ behaves like $\mathcal{O}(n^{-2})$ as $n\to\infty$.
 \end{itemize}

As the poles occur in pairs in the momentum plane we restrict
ourselves to those in the fourth quadrant. To prove the claims
made we are going to compute explicitly the functions
$\Phi^{(i)}_{k,\,l}$ and to expand then the equation
(\ref{eq:pol.eq}) in terms of $k^{-1}$. The computation is tedious
but straightforward; it yields
 \begin{eqnarray}
 \det \lambda &=& e^{2ikR}\left\{ 2kR\,L_{-1} + L_0 +
 \frac{L_1}{2kR} + \frac{L_2}{(2kR)^2} + \O(k^{-3}) \right\} \nonumber \\
 && + 2kR\,\tilde{L}_{-1} + \tilde{L}_0 + \frac{\tilde{L}_1}{2kR} +
 \O(k^{-2})\,, \label{eq:rozvoj}
 \end{eqnarray}
where
 \begin{eqnarray*}
L_{-1} &=& -\beta\frac{i}{4R}  \nonumber \\
L_0 &=& \beta\frac{1}{R}B_{-1} - 2 \re \gamma C_{-1} \nonumber \\
L_1 &=& \beta\frac{i}{R}B_{0} - 2i\re \gamma C_{0} - \alpha i D_1 \nonumber \\
L_2 &=& \beta\frac{1}{R}B_{1} - 2 \re \gamma C_{1} - \alpha D_2 \nonumber \\
\tilde{L}_{-1} &=& \beta\frac{i}{4R} \nonumber \\
\tilde{L}_0 &=& -1 -\frac{1}{4}(\alpha\beta + |\gamma|^2) \nonumber \\
\tilde{L}_1 &=& \beta\frac{i}{R}\tilde{B}_0 - \alpha i \tilde{D}_1
 \end{eqnarray*}
and the real constants $B,\,C,\,D$ , which we do not present
explicitly, depends only on the partial-wave index $l$. Three
distinct cases naturally arise: \\ [.3em]
\emph{$\delta$-type interaction:} If $\beta = \gamma = 0$ (general
$\delta$-type interaction with $\im \gamma \neq 0$, is unitarily
equivalent to this case with $\alpha'=\frac{4 \alpha}{\im^2\gamma
+ 4}$) it holds $L_{-1} = L_0 = 0$ and the resolvent-pole equation
(\ref{eq:pol.eq}) takes the form
 $$
 1 + \frac{i\alpha}{2k} = \frac{i\alpha}{2k}(-1)^l\,
  e^{2ikR}\left(1 + i\frac{l(l+1)}{kR}\right)+ \O(k^{-2})
 $$
In the leading order we have $1 = \frac{i\alpha}{2k}(-1)^l\,
e^{2ikR}$. Taking the absolute value we see that $\im k = o(k)$ as
$|k|\to\infty$ so we obtain $\frac{2 \re k}{|\alpha |} = e^{-2R\,
\im k}$ up to higher order terms; substituting back to the first
equation we get the asymptotic formulae
 $$
 k_n = \left\{ \begin{array}{lcl}
 \frac{1}{2R}\left(2n\pi +  l\pi + \frac{3\pi}{2}\right)
 + O(n^{-1}\ln n) & \mathrm{for} & \alpha > 0 \\ [.3em]
 \frac{1}{2R}\left(2n\pi +  l\pi + \frac{\pi}{2}\right)
 + O(n^{-1}\ln n) & \mathrm{for} & \alpha < 0
 \end{array} \right.
 $$
and
 $$
 \im k_n = -\frac{1}{2R} \ln \frac{2|\re k_n|}{|\alpha|} \, \left(
 1+\O(n^{-1}) \right)\,.
 $$

\noindent\emph{Intermediate-type interaction:} If $\beta = 0 $ and
$ \re \gamma \neq 0 $ the resolvent-pole equation
(\ref{eq:pol.eq}) takes the form
 $$
 e^{2ikR} \left(L_0 + \frac{L_1}{2kR} + \frac{L_2}{(2kR)^2} \right)
+ \tilde{L}_0 + \frac{\tilde{L}_1}{2kR} = \O(k^{-2})\,,
 $$
so in the leading order of $ k^{-1} $ we have
$$ L_0 e^{2ikR} + \tilde{L}_0 = 0 $$
which shows that $\im k$ is asymptotically constant because both
coefficients are nonzero. Using the explicit forms of $L_i$ for
$\beta=0$ we get
$$ k_n = \left\{ \begin{array}{l}
-\frac{i}{2R} \ln \left(\frac{1 + 1/4|\gamma|^2}{\re \gamma}
\right) + \frac{1}{R}\left( \pi n+ \frac{\pi l}{2} +\frac{\pi}{2}
\right) + \O(n^{-1})  \\ [.3em] -\frac{i}{2R} \ln \left(\frac{1 +
1/4|\gamma|^2}{-\re \gamma} \right) + \frac{1}{R}\left( \pi n+
\frac{\pi l}{2} +\frac{3 \pi}{2} \right) +
                    \O(n^{-1})
                 \end{array} \right. $$
where the upper expression holds for $\re \gamma >0$ and the lower
one for $\re \gamma < 0$. \\ [.3em]
\emph{$\delta'$-type interaction:} In the most general case when
$\beta \neq 0$ the same is true for $L_{-1}$, and therefore in the
leading order of $k^{-1}$ we get the equation
$$ 2kR\,L_{-1} e^{2ikR} + 2kR\, \tilde{L}_{-1} = 0.$$
Its solutions,
$$ k_n^{(0)} = \frac{\pi n}{R} + \frac{\pi (l+1)}{2R}\,, $$
are real which implies that $\im k_n \to 0$ as $n\to\infty$. To
obtain the convergence rate, we have to expand the expression
(\ref{eq:rozvoj}) in the vicinity of $k_n^{(0)}$. This shows that
the first nontrivial contribution to the imaginary part of
resonance is of the second order in $(k_n^{(0)})^{-1}$, hence we
deal with a quadratic equation which yields the asymptotic formula
 \begin{eqnarray*}
k_n &=& k_n^{(0)} - \frac{1}{k_n^{(0)}}\! \left(
\frac{1}{R^2}\frac{l^2 + l}{2} + \frac{1}{\beta R}\left(\re \gamma
-1 -\frac{1}{4}\abg \right) \right)  \\  && - \frac{i}{(\beta R
k_n^{(0)})^2}\! \left(1 + \frac{1}{2}|\gamma|^2 -(\re \gamma)^2
-\frac{\alpha \beta}{2} + \frac{1}{16}\abg^2 \right)  + \O(n^{-3})
 \end{eqnarray*}
This finishes the proof of the above claims.

In order to suppress additional indices  we restrict ourselves to
$\mathbb{R}^3$, but the above formulae holds for general $n$ with
simple substitution $l \to \nu -1/2$.

\begin{figure}
  \includegraphics{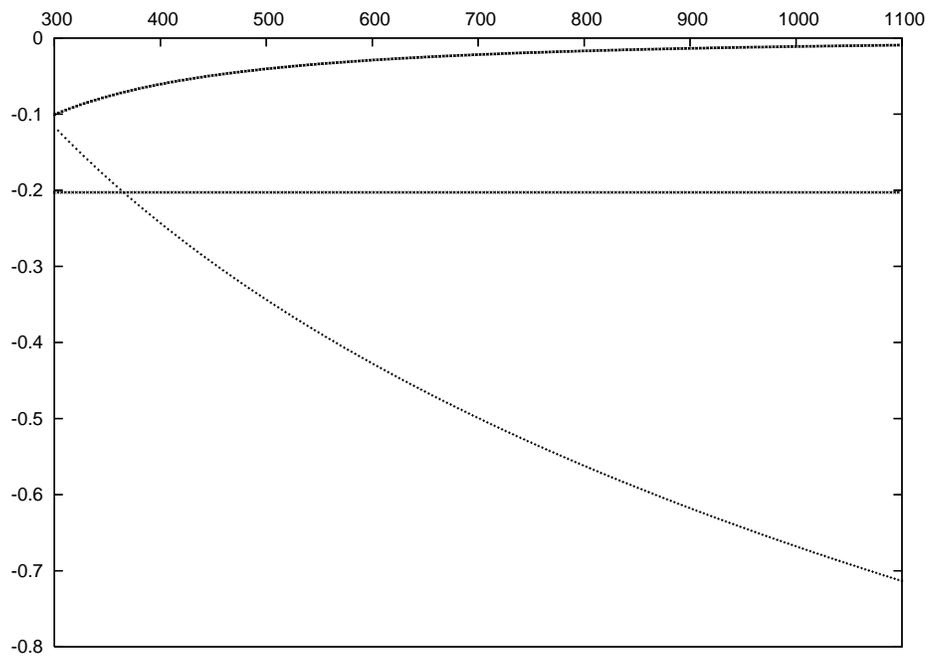}\\
  \caption{Asymptotics of the resonances in the momentum plane. Here
  the $\delta$-type poles are marked by $+$, the intermediate ones by
  $\times$ and the $\delta'$-type ones by $\ast$. }\label{fig:poles}
\end{figure}

To illustrate this result, we present in Fig.~\ref{fig:poles} the
pole behaviour for a sphere of radius $R=1$ in the three cases. To
show them on the single chart we choose the parameters $\alpha =
50,\,\gamma=\beta=0,\,$ for the $\delta$-interaction, $\alpha =
\beta= 0,\,\gamma=1 + i,\,$ for the intermediate interaction and
$\alpha = \gamma=0,\,\beta=0.01,\,$ for the $\delta'$-interaction.

\end{document}